\null
%\nopagenumbers
\magnification=\magstep1
\hsize=16.2truecm
\vsize=23.5truecm
\voffset=0\baselineskip
\parindent=1truecm
\tolerance=10000
\baselineskip=17pt
\voffset=0\baselineskip
\parskip=0.5truecm
%\par
 
\def\al{\alpha}
\def\be{\beta}
\def\ga{\gamma}
\def\c{\nabla}

\def\d{\partial}

\def\de{\delta}
\def\e{\epsilon}
\def\f{\phi}
\def\De{\Delta}

\def\la{\lambda}
\def\La{\Lambda}

\def\oo{\over}

\def\si{\sigma}

\def\ta{\tau}
\def\~{\tilde}
\def\^{\hat}
\font\twentyninerm   = lcmss8  scaled 3583 % roman
\noindent
{\bf SOLUTION OF THE VACUUM KERR-SCHILD
PROBLEM}\footnote{$^\dagger$}{Research
supported by OTKA fund no. 1826}\hfill
\vskip .1in
\noindent
{\sl L\'aszl\'o \'A. Gergely and Zolt\'an Perj\'es}\hfil
\vskip .1in
\noindent
\line{KFKI Institute for Particle and Nuclear Physics\hfil}
\line{H-1525 Budapest 114, P.O.Box 49, Hungary\hfil}
\vskip.2in
\noindent
{\bf ABSTRACT}
\midinsert
\baselineskip=12pt plus.1pt
   {\sl The complete solution of the vacuum Kerr-Schild equations
in general relativity is presented, including the space-times with a curved
background metric. The corresponding result for a flat background
has been obtained by Kerr.}
 
\endinsert
\baselineskip=17pt
\voffset=0\baselineskip
\par
 
\smallskip
 
  The Kerr-Schild pencils$^1$ of metrics
$$\tilde g_{ab}=g_{ab}+V l_a l_b          \eqno(1)$$
have been in the forefront of research in general relativity
for some time. A classic example of these space-times is the
Kerr metric. All the Kerr-Schild vacuum space-times with a flat
parent metric $g_{ab}$ are given by
beautiful geometrical relations$^2$.
The solution of the flat problem (known as {\it Kerr's theorem})
reveals a close relationship with complex
surfaces in three-dimensional homogeneous spaces.
Kerr-Schild congruences in Minkowski space-time have extensively
been studied$^{3-6}$.
Work on Kerr-Schild space-times in the generic case
when $g_{ab}$ has a nonvanishing curvature, $l$ is a null vector
and $V$ a function, has certainly been motivated by the prospects
of extending Kerr's complex analytic description to curved
space-times$^{7-10}$.
 
   In this Letter we present the complete solution of the
vacuum Kerr-Schild problem.
The Kerr-Schild equations follow from (1) and the vacuum Einstein
equations. They are:
$$Dl_a=0  \ ,                                        \eqno(2)$$
$$\c_b\bigl[\c_a\left(Vl_c l^b\right)+\c_c\left(Vl_a
l^b\right)-\c^b\left(Vl_a l_c\right)\bigr]=0   \eqno(3)$$
$$DDV+(\c^a l_a)DV+2V(\c_b l_d)\c^{[b}l^{d]}=0.\eqno(4)$$
\vfill\eject
Here $\c$ is the covariant derivative annihilating $g_{ab}$.
The vector
  $$D=l^a\c_a=\partial/\partial r   \eqno(5)$$
is tangent to a null geodesic congruence$^8$, with $r$ the
affine parameter.
 
   We rewrite these equations in a Newman-Penrose (NP)
notation$^{12}$, choosing $l$ a vector of the null tetrad.
The geodesic condition (2) becomes $ \kappa=0$.
We adopt the gauge with
$\e=0\ ,\qquad  \pi=\al+\bar\be $,
and integrate a closed subset of the
`radial' field equations$^{12}$ for
the affine parameter dependence:
$$\rho=-{1\oo 2r}(1+\cos\eta\ C)    \ ,     \qquad
  \si=-{\sin\eta\oo 2rC}\                \eqno(6)$$
$$\Psi_0=-{\sin2\eta\oo 4r^2}        \ . \eqno(7)$$
Here C is a complex phase factor
$$C={r^{\cos\eta}-iB\oo r^{\cos\eta}+iB} \ .\eqno(8)$$
$B$ and $\eta$ are real integration functions.
A further integration function has been
eliminated by suitably fixing the origin of
the affine parameter. The real potential can be written as
$$V= V_0{r^{\cos\eta}\oo r^{2\cos\eta}+B^2}\ .\eqno (9)$$
The complex tetrad vector $m$ has the $r$ dependence
$$m={1\oo2B}\Bigl(1-{1\oo C}\Bigr)
\Bigl[iQ_1^j r^{\cos\eta-\sin\eta-1\oo2}-Q_2^j r^{\cos\eta+\sin\eta-1\oo2}
\Bigr]\,{\d\oo\d x^j}\qquad j=1,2,3 \ \eqno(10)$$
where $V_0,Q_1^j$ and $Q_2^j$ are real integration functions.
With this choice, the tetrad has been uniquely fixed.
 
  The real function $B$ controls the amount of divergence
and rotation of the null congruence with tangent $l$, respectively.
$B$ gives rise to the
imaginary part of the spin coefficient $\rho$. When
$B=0$, we have $C=1$, and the congruence
is curl-free. Similarly, for large values of the affine parameter
$r$, the phase factor $C$ approaches the unit value, and the
rotation dies out. The parameter $\eta$ governs the shear.
For $\eta=0$ or $\eta=180^\circ$, the congruence
is shear-free. When both $B=0$ and
$\eta=0$, the rays are exactly spherical: $\rho=-1/r$.
For $\eta=90^\circ$, the rays become cylindrical, $\rho=-1/2r$.
At $\eta=180^\circ$, there is no expansion.
Our Theorem below shows that the general shearing class
does not contain the shear-free case as a smooth limit.
 
   The field equations in the NP form may
be grouped in three sets. The first set of equations is a
coupled system of linear inhomogeneous equations for the affine
parameter dependence of the quantities $\ta,\pi,\al,\be,\Psi_1$
and for their complex conjugates:
$$\eqalign{\twentyninerm T&HE\cr&\Psi_1\cr
E&QS.\cr}\qquad\qquad\qquad\qquad \eqalign{
D\ta&=\rho(\ta+\bar\pi)+\si(\bar\ta+\pi)+\Psi_1  \cr
D\pi&=2\rho\pi+2\bar\si\bar\pi+\bar\Psi_1        \cr
D\al&=\rho(\pi+\al)+\bar\si(\bar\pi-\bar\al)     \cr
D\be&=\bar\rho\be+\si(2\pi-\bar\be)+\Psi_1       \cr
D\Psi_1&=4\rho\Psi_1+(\bar\de+\pi-4\al)\Psi_0.   \cr}
\qquad\qquad\qquad\qquad\eqalign{
&(11a) \cr
&(11b) \cr
&(11c) \cr
&(11d) \cr
&(11e) \cr}$$
The $\Psi_1$ equations are
complemented by two equations, linear in the unknown functions:
$$\de\Bigl({\Psi_0\oo2\si}\Bigr)+{\Psi_0\oo\si}\de
ln\phi-2\si\bar\de ln\phi-
\Psi_1={\Psi_0\oo2\si}(\ta-\bar\al-\be)-2\si(\bar\ta-\al-\bar\be)
+\bar\ta\si-\ta\rho                              \eqno(12)$$
$$\de\rho-\bar\de\si=\rho\bar\pi+\si(\pi-4\al)+(\rho-\bar\rho)\ta-\Psi_1
                                                 \ .\eqno(13)$$
where $\phi=\sqrt V$.
 
  \item{ }{\bf Theorem:} {\sl For a generalized vacuum-vacuum
Kerr-Schild pencil, either of the following conditions holds:
The parameter $\eta$ assumes one of the special values given by
$$\sin\eta=0,\ ^+_-1,\ ^+_-2^{-{1\oo 2}}\quad \ .\eqno(14)$$
or, alternatively, the spin coefficient quantities
$\rho,\si$ and $\Psi_0$ depend only on the affine parameter $r$.}
 
   The theorem can be proven$^{11}$ by taking the $D$ derivative
of Eq. (12), and eliminating the unknown spin coefficients by use
of the NP commutators, Eq. (13) and the equations of the $\Psi_1$ system.
This yields that $\de\eta=0$ and, unless $\eta$ takes any of the
exceptional values, the integration functions are restricted by
$\de V_0=\de B=0$. In the generic case, $\rho-\bar\rho\neq0$,
the commutator equations imply also $\De\f=0$
for any real function $\f$ with $D\f=\de\f=0$,
Hence the integration functions $B,\eta$ and $V_0$ are
constants.
The curl-free fields, $B=0$, will be discussed in Ref. 11.
 
 It follows that the Kerr solution with $\eta=0$
will {\sl not} emerge as a smooth limiting case of the shearing
Kerr-Schild metrics.
 
    For fields of arbitrary deformation parameter,
$\al$ and $\Psi_1$ may be expressed algebraically
in terms of $\pi,\ta$ and their complex conjugates,
by the respective Eqs. (12) and (13).
The $\Psi_1$ system becomes a quartet of coupled
equations for the spin coefficients $\ta,\pi$ and their complex
conjugates:
$$\eqalign{&
D\ta=(2\rho-{\Psi_0\oo2\si})\ta+\si(2\bar\ta-\pi)
                +(\rho+{\Psi_0\oo2\si})\bar\pi\cr
& D\pi=(2\rho+{\Psi_0\oo2\bar\si})\pi+\bar\si\ta
        + (\bar\rho-{\Psi_0\oo2\bar\si})\bar\ta} \eqno(15)  $$
along with the complex conjugate equations.
The fundamental solution of the $\Psi_1$ set is given in Table~1.
The general solution is a linear combination of the four
fundamental solution vectors with {\sl real} coefficients.
 
  The radial component of the vector $n$ may be obtained by
applying the commutator $[\de,\bar\de]$ on r. Hence the operator $\De$,
when acting on any of the functions depending only on $r$, takes the form
$\De= -{\bar\mu-\mu\oo\bar\rho-\rho}D$. The second set of field
equations consists
of the NP Eqs. (4.2l,p,q), as well as the fifth of NP (4.5) and of
the Kerr-Schild equation
$$\eqalignno{
\Bigl[{1\oo2}&{\bar\rho+\rho\oo\bar\rho-\rho}\Bigl({1\oo
      r}+\bar\rho+\rho\Bigr)\Bigr](\bar\mu-\mu)
      +{1\oo2}\Bigl({\Psi_0\oo\si}
        +\rho-\bar\rho\Bigr)(\bar\mu+\mu)
     +\Psi_2+\bar\Psi_2-(\rho+\bar\rho)(\ga+\bar\ga) \cr
  =&\de(\bar\ta-\pi)+\bar\de(\ta-\bar\pi)-6\pi\bar\pi+2\pi\bar\al
+2\bar\pi\al-2\ta\bar\ta-2\ta\al-2\bar\ta\bar\al
+3\ta\pi+3\bar\ta\bar\pi  \ .
                                              &(16)
}$$
We obtain a lengthy relation from these, containing only functions
the $r$-dependence of which is explicitly known. Collecting
the coefficients of independent powers of $r$, one finds that
nothing but the trivial solution of the $\Psi_1$ system satisfies
this equation:
$$\al=\be=\ta=\pi=\Psi_1=0     \ .\eqno(17)$$
Thus the second system of
equations is homogeneous and linear in
$\mu,\la,\ga$ and $\Psi_2$.  The determinant vanishes, and we get
$${\ga\oo\rho+1/2r}={\la\oo\bar\si}=-{\Psi_2\si\oo\Psi_0\rho}={\mu\oo\rho}
                          \ . \eqno(18)$$
The remaining field equations constitute the third set.
They further restrict the metric, leaving us with a
three-parameter pencil for which the image
of the Kerr-Schild map is the K\'ota-Perj\'es$^{13}$ metric (44):
$$ds^2=-{r^{\cos2\eta}+B^2\oo r^{\cos\eta}}
  (dr^2+r^{1-\sin\eta}dx^2+r^{1+\sin\eta}dy^2)+
  \La{r^{\cos\eta}\oo r^{\cos2\eta}+B^2}(l_adx^a)^2\ .\eqno(19)$$
Here $l=\partial/\partial r$ is tangent to the Kerr-Schild
congruence and $\La$ is the pencil parameter.
 
   To complete our investigation of vacuum
Kerr-Schild space-times, we consider now in turn the metrics with
either of the values (14).
 
   (a) When $\sin\eta=0$, both $\si$ and $\Psi_0$ vanish,
and $l$ is a principal null vector of the curvature. By the
Goldberg-Sachs theorem$^{12}$, these parent space-times are
algebraically special, and $\Psi_1=0$.
It then follows from Thompson's Theorem 3.2
that also the ensuing space-time is algebraically special, with
the Kerr-Schild congruence being a principal null congruence$^8$.
All the vacuum Kerr-Schild spacetimes generated
from the flat space-time are in this class.
 
   (b) The case with $\cos\eta=0$ maps Minkowski space-time
to itself.
 
   (c) Case $\sin\eta=1/\sqrt{2}=k$ contains the following
K\'ota-Perj\'es metrics:
$$ds^2=-{f^0\oo f}(r^{1-k}dx^2+r^{1+k}dy^2)+2dr
(l_a dx^a)+f(l_a dx^a)^2 \eqno(20)\ .$$
Metric (53) of Ref. 8 is given by
$$f=\La Re\Bigl\{{x+ir^ky\oo r^k+iB}\Bigr\}, \qquad f^0=\La(x+By)
\eqno(21)$$
with $B$ a real constant.
For metric (66), $B=x/y$ and
$$f=\La {x+by\oo x^2r^{-k}+y^2 r^k}, \qquad f^0=\La(x+by)/y^2\ .\eqno(22)$$
Notice that we have enlisted above all the metrics in Ref. 8.
 
  We thus find that all K\'ota-Perj\'es metrics are explicit
cases of Kerr-Schild space-times, either with a real deformation
parameter or with $\eta=1/\sqrt2$.
 
   In summary, the structure of our solution is
as follows. The Kerr-Schild space-times are
characterized by the real deformation parameter $\eta$.
The deformation parameter vanishes for Kerr-Schild space-times
the parent of which is Minkowski space-time. By our theorem,
the field quantities are severely restricted unless the
parameter $\eta$ assumes either of the exceptional values
$0,\pm1, \pm\sqrt2/2$. The metrics
with arbitrary values of the deformation parameter are
K\'ota-Perj\'es metrics. For the exceptional values of $\eta$,
(a) the class with $\eta=0$ is algebraically
special, (b) the values $\sin\eta=\pm1$ can occur only in
automorphisms of the Minkowski space-time, and (c) the class with
$\sin\eta=\pm1/\sqrt2$ contains the remaining K\'ota-Perj\'es metrics.
While the Kerr-Schild congruences in a Minkowski space-time form
a four-parameter family$^2$, there is no corresponding structure
on a curved background. Our results have the important implication
that hopes are dashed for a complex-analytic description of space-time within
the framework of Kerr-Schild theory.
 {\ \ }
\bigskip
{\bf REFERENCES}
{\frenchspacing
\medskip
\item{[1]} Kerr, R. P.  and Schild, A., {\it Atti Del
Convegno Sulla Relativit  Generale: Problemi Dell' Energia e Onde
Gravitazionali (Anniversary Volume, Fourth Centenary of Galileo's
Birth)}, G. Barb'ra, Ed. (Firenze, 1965), p. 173
\item{[2]} Debney, G. C., Kerr, R. P.  and Schild,
A., J. Math. Phys. {\bf10}, 1842 (1969)
\item{[3]} Boyer, R. H. and Lindquist. R. W.,
J. Math. Phys. {\bf8}, 256 (1967)
\item{[4]} Grses, M. and Grsey, F.
J. Math. Phys. {\bf16}, 2385 (1975)
\item{[5]} Debever, R. , Bull. Acad. Roy. Belgique Cl. Sci.
{\bf60}, 998 (1974)
\item{[6]} C. B.G.McIntosh: Kerr-Schild Spacetimes Revisited,
in {\it Conference on Mathematical Relativity}, Ed. R. Bartnik,
Proc. Centre for Mathematical Analysis,
Australian National University, Canberra, 1988.
\item{[7]} D. Kramer {\it et al.}: {\it Exact Solutions of
Einstein Field Equations}, Cambridge Univ. Press (1980)
\item{[8]} Thompson, A.H., Tensor {\bf17}, 92 (1966)
\item{[9]} Dozmorov, I.M., Izv. VUZ. Fiz.{\bf11}, 68 (1971)
\item{[10]} E. Nahmad-Achar, J.Math.Phys. {\bf29},1878 (1988)
\item{[11]} Gergely, L. \'A. and Perj\'es, Z.: Kerr-Schild metrics
revisited I.-II., submitted for publication
\item{[12]} Newman, E. and Penrose, R., J. Math. Phys. {\bf3}, 566
(1962)
\item{[13]} K\'ota, J. and Perj\'es, Z., J. Math. Phys. {\bf13}, 1695
(1972)
 
\vfill\eject}
 
\par
\smallskip

\def\c{\cos\eta}
\newbox\ujstrutbox
$$\eqalign{
\pi^{(1)}=\ {C+1\oo r^{\cos\eta+\sin\eta+1\oo2}}
&\Bigl(C^{-3}-{\cos\eta\oo \sin\eta+1}C^{-2}
 -{\sin\eta\oo \sin\eta+1}C^{-1}                     \cr
&+{5sin^2\eta-4\sin\eta+3\oo \cos\eta(\sin\eta-3)}
+{3\sin\eta-1\oo \sin\eta+1}{2\sin\eta+3\oo \sin\eta-3}C
+2{\sin\eta\oo \cos\eta}C^2\Bigr)                    \cr\cr
\pi^{(2)}=i{C+1\oo r^{\cos\eta-\sin\eta+1\oo2}}
&\Bigl(C^{-3}+{\cos\eta\oo \sin\eta-1}C^{-2}
-{\sin\eta\oo \sin\eta-1}C^{-1}                      \cr
&+{5cos^2\eta-4\sin\eta-8\oo \cos\eta(\sin\eta+3)}
+{6cos^2\eta+7\sin\eta-3\oo cos^2\eta-2 \sin\eta+2}C
-2{\sin\eta\oo \cos\eta}C^2\Bigr)                    \cr\cr
\pi^{(3)}=i{C-1\oo r^{\cos\eta+\sin\eta+3\oo2}}
&(C+1)^2\Bigl(-C^{-3}-{\sin\eta+1\oo \cos\eta}C^{-2}
+{1\oo \sin\eta-1}C^{-1}
-2{\sin\eta\oo \cos\eta}\Bigr)                       \cr\cr
\pi^{(4)}=\ {C-1\oo r^{\cos\eta-\sin\eta+3\oo2}}
&(C+1)^2\Bigl(-C^{-3}-{\cos\eta\oo \sin\eta+1}C^{-2}
-{1\oo \sin\eta+1}C^{-1}
+2{\sin\eta\oo \cos\eta}\Bigr)                       \cr\cr
                                    }$$
$$\eqalign{
\tau^{(1)}=\ {C+1\oo r^{\cos\eta+\sin\eta+1\oo2}}
&\Bigl({\sin\eta\oo \cos\eta}C^{-3}-{\sin\eta\oo \sin\eta+1}C^{-2}
 -{sin^2\eta+8\sin\eta+3\oo \cos\eta(\sin\eta-3)}{\sin\eta-1\oo
\sin\eta+1}C^{-1}   \cr
&-3{\sin\eta\oo \sin\eta+1}
+{2\sin\eta-3\oo \cos\eta}C
+2C^2\Bigr)                    \cr\cr
\tau^{(2)}=i{C+1\oo r^{\cos\eta-\sin\eta+1\oo2}}
&\Bigl({\sin\eta\oo \cos\eta}C^{-3}+{\sin\eta\oo \sin\eta-1}C^{-2}
 -{\sin\eta cos^2\eta-7cos^2\eta+4\sin\eta+4\oo
\cos\eta(2\sin\eta-2-cos^2\eta)}C^{-1}   \cr
&+3{\sin\eta\oo \sin\eta-1}
+{2\sin\eta+3\oo \cos\eta}C
-2C^2\Bigr)                    \cr\cr
\tau^{(3)}=i{C-1\oo r^{\cos\eta+\sin\eta+3\oo2}}
&(C+1)^2\Bigl(-{\sin\eta\oo \cos\eta}C^{-3}+{\sin\eta\oo
\sin\eta-1}C^{-2} -{2\sin\eta+1\oo \cos\eta}C^{-1}
-2\Bigr)                       \cr\cr
\tau^{(4)}=\ {C-1\oo r^{\cos\eta-\sin\eta+3\oo2}}
&(C+1)^2\Bigl(-{\sin\eta\oo \cos\eta}C^{-3}-{\sin\eta\oo
\sin\eta+1}C^{-2} -{2\sin\eta-1\oo \cos\eta}C^{-1}
+2\Bigr)                 \ .      \cr
                                    }$$
\par
\smallskip
\ \ \ \ \ Table 1. {\it The four solutions for $\pi$ and
$\ta$} \par
\smallskip
\par
\end